\documentclass[pra,twocolumn,groupedaddress]{revtex4-1} % superscriptaddress
\usepackage[T1]{fontenc}
\usepackage[english]{babel}
\usepackage{siunitx}
\usepackage[utf8]{inputenc}
\usepackage{url}
\usepackage{dsfont}
\usepackage{bbm}
\usepackage{nicefrac}
\usepackage{setspace}
\usepackage[short]{optidef}
\usepackage{float}
\usepackage{comment}
\usepackage{dsfont}
\usepackage{amssymb}  
\usepackage[colorlinks=true, citecolor=blue,urlcolor=blue,linkcolor=blue,filecolor=black]{hyperref}
\usepackage[colorinlistoftodos, color=green!40, prependcaption]{todonotes}
\usepackage{ulem}
\usepackage{amsthm}
\usepackage{mathtools}
\usepackage{physics}
\usepackage{graphicx}
\usepackage{adjustbox}
\usepackage{placeins}
\usepackage{lipsum}
\usepackage{csquotes}
\usepackage{blindtext}

\newcolumntype{"}{@{\hskip\tabcolsep\vrule width 1pt\hskip\tabcolsep}}

\newcommand{\beq}{\begin{equation}}
\newcommand{\eeq}{\end{equation}}
\renewcommand{\emph}{\textit}

\usepackage{soul}
\usepackage{graphicx}% Include figure files
\usepackage{dcolumn}% Align table columns on decimal point
\usepackage{bm}% bold math

\usepackage[utf8]{inputenc}
\usepackage[T1]{fontenc}
\usepackage{mathptmx}

\begin{document}

\preprint{AIP/123-QED}

\title[]{Photonic integrated chip enabling orbital angular momentum multiplexing for quantum communication}%Photonic integrated chip excitation of orbital angular momentum fiber modes for quantum communication
% Force line breaks with \\

\author{Mujtaba Zahidy} \author{Yaoxin Liu} \author{Daniele Cozzolino} \author{Yunhong Ding} \author{Toshio Morioka} \author{Leif K. Oxenløwe} \author{Davide Bacco}\email{dabac@fotonik.dtu.dk}
\affiliation{Center for Silicon Photonics for Optical Communications (SPOC), Department of Photonics Engineering, Technical University of Denmark, Kgs. Lyngby, Denmark}


\begin{abstract}\noindent
%The demand for higher key rate in quantum key distribution applications necessitates novel approaches for overcoming state-of-the-art limitation imposed by current technology. Multiplexing quantum signals exploiting different degrees (e.g., wavelength, space, polarization) of freedom is a common approach to increase the capacity of the channel. Here, we show how a photonic integrated chip, designed to excite orbital angular momentum modes in a ring-core fiber, can be used for spatial multiplexing, allowing to perform parallel QKD in 3 different modes over a 800 m long fiber. The experiment sets the first steps towards quantum OAM division multiplexing enabled by a compact and light-weight silicon chip and further push the development of integrated scalable devices supporting orbital angular momentum modes. 
%
Light carrying orbital angular momentum constitutes an important resource for both classical and quantum information technologies. Its inherently unbounded nature can be exploited to generate high-dimensional quantum states or for channel multiplexing in classical and quantum communication in order to significantly boost the data capacity and the secret key rate, respectively. While the big potentials of light owning orbital angular momentum have been widely ascertained, its technological deployment is still limited by the difficulties deriving from the fabrication of integrated and scalable photonic devices able to generate and manipulate it. Here, we present a photonic integrated chip able to excite orbital angular momentum modes in an 800 m long ring-core fiber, allowing us to perform parallel quantum key distribution using 2 and 3 different modes simultaneously. The experiment sets the first steps towards quantum orbital angular momentum division multiplexing enabled by a compact and light-weight silicon chip, and further pushes the development of integrated scalable devices supporting orbital angular momentum modes.         
\end{abstract}
\maketitle

\section{Introduction}
%Intro about the OAM and its features
Since 1992, when L. Allen and colleagues discovered that Laguerre-Gaussian beams have a well-defined orbital angular momentum (OAM) \cite{allen1992orbital}, an enormous amount of research has been carried out to better understand and manipulate light owning a nonzero OAM \cite{shen2019optical}. Optical beams possessing a well-defined OAM are characterized by the azimuthal phase dependence $e^{i\ell\varphi}$, where $\ell\hbar$ is the of OAM carried by each photon, with $\hbar$ being the reduced Planck constant; $\ell$ is the topological charge, an integer that specifies the OAM value, and $\varphi$ is the azimuthal angle. Such helical phase twists along its propagation axis and determines the cancellation of the light beam at the axis itself, thus resulting in a “doughnut” intensity profile. It is thanks to these special features, i.e., intensity and phase structures, that light owning an OAM has been applied in many fields of optics, e.g. optical trapping \cite{paterson2001controlled,padgett2011tweezers} or quantum information \cite{erhard2018twisted,forbes2019quantum}.
%OAM for communication
In the last decade, OAM has been largely investigated in the field of fiber-based optical communication, both classical and quantum, achieving unprecedented results that have forecast its exploitation to real-case scenarios \cite{willner2015optical,cozzolino2019orbital,wang2021high}. It has shown great potentials in communication systems due to the unbounded nature of the topological charge $\ell$ and the inherent orthogonality between optical modes or quantum states. Indeed, these characteristics are exceptional resources for optical mode multiplexing and high-dimensional quantum communication: the former aims to overcome the channel capacity crunch in classical communication systems \cite{bozinovic2013terabit,wang2012terabit} or to boost photon information efficiency in the quantum ones \cite{cozzolino2019orbital}; the latter, uses quantum states encoded in a large Hilbert space, i.e. high-dimensional quantum states, as they can tolerate higher noise thresholds, thus they are useful for communications over noisy channels or in extreme conditions, such as photon starving or detector saturation regimes \cite{cozzolino2019high,ecker2019overcoming}.  
%Need for integration/integrated devices
Nonetheless, despite the results achieved hitherto, applications of OAM beyond proof-of-principle experiments require the development of integrated devices able to generate, transmit and manipulate such a degree of freedom. On-chip generation of OAM modes both for classical and quantum applications, have been demonstrated using star couplers \cite{doerr2011silicon}, microring resonators \cite{cai2012integrated}, and controlled phase arrays \cite{sun2014generating}. Also, the transmission of OAM modes through a silica chip has been investigated \cite{chen2018mapping}, as well as the combination of an integrated optical emitter and a ring-core fiber for classical communication \cite{liu2018direct}.\\   
%In this paper...
In this work, we exploit a photonic integrated emitter based on the star-coupler technology to seed a 800 m long ring-core fiber with a three-times OAM multiplexed quantum key distribution (QKD) protocol, using time-bin encoded states. We believe our work further closes the gap between proof-of-principle experiments and concrete deployment of OAM-based technologies, thus foreseeing them as a near-future reach.   

\section{Excitation of the orbital angular momentum fiber modes}
\begin{figure*}[ht]
\includegraphics[width=\textwidth]{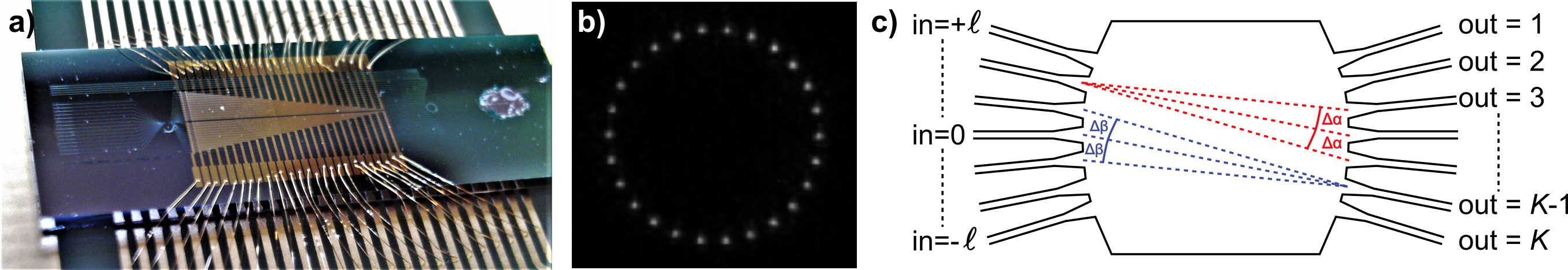}
\caption{a) Picture shows the photonic integrated device used in the experiment. b) Infrared image of the grating couplers output. c) Schematic of the star-coupler structure.}
\label{fig:chip}
\centering
\end{figure*}

The main goal of our experiment is to spatially multiplex and transmit time-bin encoded QKD signals using the OAM degree of freedom. In this regard, a fundamental role is played by our integrated device, which is a silicon-on-insulator chip that, given an input, outputs a ring of Gaussian spots with a well-defined relative phase \cite{Baumann19}. Figure \ref{fig:chip} a) and b) shows the fabricated device and its output. The chip consists of three main parts: the input grating couplers, a star coupler and a ring of output couplers, and it supports the multiplexing to OAM modes within $\ell=-7$ to $\ell=+7$. There are 15 input grating couplers with a 500 $\mu$m adiabatic taper, whose spacing is 127 $\mu$m that is compatible with commercially available fiber arrays. Each input can be labelled with integers ranging from $-\ell$ to $+\ell$. 
% --- Figure Chip
% --- End figure
The star coupler \cite{doerr2011silicon}, schematically shown in Figure \ref{fig:chip} c), is an optical element that distributes an incoming signal into $K$ output waveguides with phase differences of $\Delta\varphi = 2\pi\ell/K$ for neighboring output waveguides, so that the total phase difference across all output waveguides is $2\pi\ell$, where $\ell$ depends on the chosen input waveguide. In our case, $\ell$ lies within -7 and +7 and $K=26$. The input and output waveguides are spread at equidistant angles $\Delta\alpha$ and $\Delta\beta$, respectively. 
The 26 output ports are connected to a ring of 26 grating couplers oriented in the same direction. Figure \ref{fig:chip} b) shows an image of the 26 grating couplers output.
The optical path length of the 26 waveguides between the star coupler and the ring of output couplers are identical, nonetheless, each waveguide is supported with a thermal heater that allows for phase compensation to account for fabrication tolerances. The chip has approximately 22 dB of losses, stemming from input coupling loss, waveguides and other components losses, and output grating couplers losses.\\
A specific OAM fiber mode $\pm\ell$ can be excited by injecting light into the respective $\pm\ell$ input, so that the output ring of Gaussian spots is generated, whose total phase difference is $2\pi\ell$. The injection of light into more inputs simultaneously enables the excitation of different OAM fiber modes, thus realizing the OAM mode multiplexing.
%The total dimension is ca. 11.7 mm x 2.3 mm.

\section{ Experimental Setup }
% --- Figure Setup
\begin{figure*}[ht]
\includegraphics[width=\textwidth]{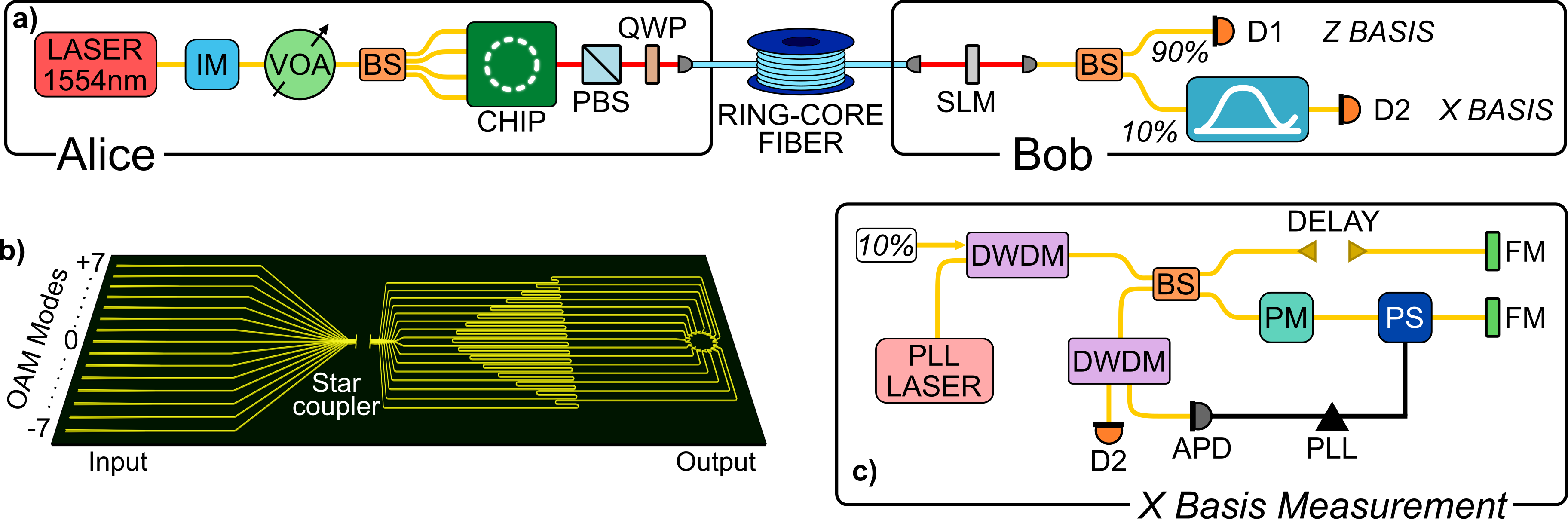}
\caption{a) Experimental setup. IM: intensity modulator; VOA: variable optical attenuator; BS: beam-splitter; PBS: polarization beam-splitter; QWP: quarter-wave plate; SLM: spatial light modulator; D$i$: i-th detector. b) Schematic of the photonic integrated device. c) $X$ basis measurement setup. PLL: phase-locked loop; DWDM: dense wavelength-division-multiplexing; PM: phase modulator; PS: phase shifter; FM: Faraday mirror; APD: avalanche photodiode; PLL: phase-locked loop.}
\label{FIG::ExpSetup}
\centering
\end{figure*}
% --- End figure
The QKD protocol we implemented and multiplexed using OAM modes is the three-state time-bin protocol, where we have used the 1-decoy technique and carried out the finite key analysis \cite{bacco2019field}. The experimental setup we realized is shown in Figure \ref{FIG::ExpSetup}. A continuous wave laser at 1554.13 nm, channel 29 of the International Telecommunication Union - Telecommunication Standardization Sector (ITU-T), is carved to form a train of pulses at a repetition rate of 595 MHz, which are subsequently attenuated to reach the single photon level and form the time-bin qubits exploited in the experiment. The carving procedure is performed by two cascaded intensity modulators, shown as one in Figure \ref{FIG::ExpSetup}, controlled by a field programmable gate array (FPGA), which generates electrical signals according to a pseudo-random binary sequence of length $l=2^{12}-1$. 
%--- Crosstalk figure
\begin{figure}[!hb]
\includegraphics[width=0.45\textwidth]{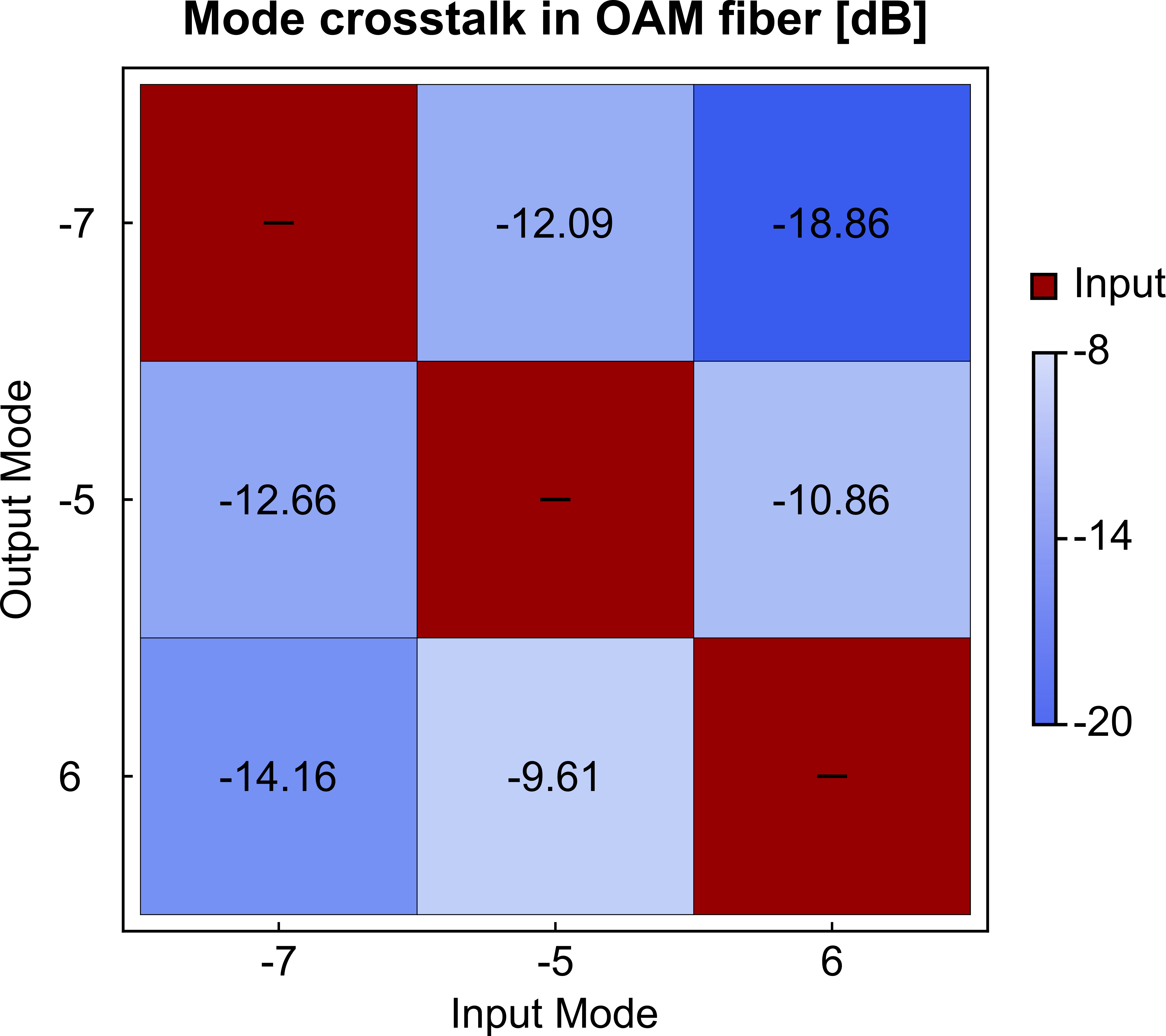}
\caption{\textbf{Mode crosstalk matrix in OAM fiber.} On the x axis we report the inputs mode (-7, -5 and 6). On the y axis we report the output mode (-7, -5 and 6). The measurement has been normalized per each mode.}
\label{FIG::CrossTalk2D}
\centering
\end{figure}
%---End figure
%
It should be noted that to block the loopholes and guarantee the security of the real QKD implementation, quantum states should be phase randomized and the pseudo-random sequence should be replaced with a quantum random number generator \cite{tebyanian2021practical}. Furthermore, in addition to providing the electrical signals driving the intensity modulators, the FPGA generates an electrical signal, of width 1.68 ns and a repetition rate of 145.358 KHz, which is used for clock synchronization. However, due to source and receiver physically situated remote, an optical synchronization scheme was implemented. Following the cascaded intensity modulators, the quantum state signal is then split by using a 1 by 4 beam-splitter and fed to the integrated chip. Since the integrated chip is polarization sensitive, polarization controllers are used to maximize coupling of individual signals into the chip. Finally, the chip output is collimated, and its polarization is transformed from linear to circular by the combination of a polarization beam-splitter and a quarter-wave plate. After that, the chip output is coupled into the ring-core fiber, corresponding to the quantum channel, exciting the desired OAM modes, each of them generating an independent key. For the purpose of this demonstration, we choose 2(3) modes, $-7$ and $-5$ ($-7$, $-5$ and $+6$) which showed a crosstalk as low as $\approx -12$ dB ($\approx -18$ dB), see Figure \ref{FIG::CrossTalk2D}.
The OAM modes crosstalk stems from misalignment of the chip output to the OAM fiber, mismatch of the chip outputs relative phase induced by the heaters, as well as bends and twists of the fiber itself. Since the group velocity of each mode in the OAM carrying fiber is different for each of them, the mode crosstalk can be directly measured and optimized with a time-of-flight measurement \cite{cozzolino2019orbital}. A further analysis of crosstalk was performed via power measurement. Exciting only one mode at a time, a spatial light modulator (SLM) is used to demultiplex the output with various modes. The power coupled to the single mode fiber is then compared with the excited mode and crosstalk is measured. In our case, a further optimization has been possible by performing multiple runs of phase optimization for the chip output \cite{YaoxinLiuECOC21}. 
At the receiver, Bob, OAM modes demultiplexing is performed by means of an SLM and one mode at a time. The Gaussian beam obtained from the SLM conversion is then coupled to a single mode fiber, which also filters out the unwanted modes residue.
%The clock synchronization signal also co-propagates in the single mode fiber with quantum signals, by means of a DWDM scheme, towards the detection unit which is located in a different laboratory. Two DWDMs filter separate the quantum signal from the clock synchronization signal where the second one filters out any remnants of the clock synchronization signal.
Following the single mode fiber coupling, a 90:10 beam-splitter marks the beginning of the passive measurement stage. The $90\%$ output, corresponding to the $Z$ basis, is detected directly with a superconducting nano-wire single photon detector (SNSPD), while the $10\%$ output, corresponding to the $X$ basis, is redirected to an unbalanced Michelson interferometer with 800 ps time delay in one arm with respect to the other one, see Figure \ref{FIG::ExpSetup} c).
The fiber-based Michelson interferometer consists of two Faraday rotator mirrors, a polarization controller, adjustable delay-line, and a piezoelectric phase shifter. To compensate for the relative phase drift of the interferometer arms, a phase lock loop is implemented \cite{da2021path}. A monitor laser (PLL laser) is mixed with the $10\%$ part of the signal via a dense wavelength-division-multiplexing (DWDM) device and sent to the interferometer. At the interferometer output, a second DWDM separates the QKD signal and the the PLL laser. By monitoring the PLL laser power with an avalanche photodiode it is then possible to actuate the phase shifter inside the interferometer and stabilize the interference fringes. 
To maximize the efficiency and the key-rate, SNSPDs were used with 83\% efficiency, $\approx$ 50 count per second of dark counts, and 33 ps of dead-time. Detection events and their time of arrival are registered by a time to digital converter with temporal resolution of 1 ps.

\section{Results}
One key parameter for QKD protocols is the quantum bit error rate (QBER). In particular, minimizing mode crosstalk is crucial for having low error rate in the QKD experiment. In this way, indeed, the error reconciliation part has less impact on the final secret key rate. In Figure \ref{FIG::CrossTalk2D}, we report the results of the mode crosstalk measurements after 6 rounds of phase optimization followed by the alignment of the chip output to the OAM fiber.
We then proceed to the implementation of the QKD experiment by sending all the modes at the same time and by demultiplexing one mode at a time. The channel loss is measured around 1 dB for 800 m of ring-core fiber, however, the signal and key rate suffered from an average 15 dB of coupling losses after the OAM-to-Gaussian mode conversion, and 9.15 dB of extra loss due to the optical synchronization scheme implemented. 
%--- 2 Modes table
\begin{table}[]
\begin{tabular}{||p{0.15\textwidth} p{0.15\textwidth} p{0.15\textwidth}|}%{||c c c|}
\hline
 & Mode 7 &  Mode 5   \\
 \hline\hline
 \rule{0pt}{2.2ex} 
 $\mu_1$ & 0.26 & 0.36  \\
 %\hline
 $\mu_2$ & 0.13 & 0.13  \\
 %\hline
 $\text{QBER}_{Z_{\mu_1}}$ & 2.15 & 1.35  \\
 %\hline
$\text{QBER}_{Z_{\mu_2}}$ & 2.13 & 2.00  \\
 %\hline
$\text{QBER}_{X_{\mu_1}}$ & 4.23 & 3.53   \\
%\hline
$\text{QBER}_{X_{\mu_2}}$ & 4.08 & 4.12   \\
%SKR [kBits/s] & 121 & 77.9  \\
\hline
\end{tabular}
\caption{Experimental mean photon numbers for the 2-mode multiplexing demonstration  as well as QBERs and SKR attained for each mode.}
\label{TAB::2OptimalMu}
\end{table}
%--- End table
%--- 3 Modes table
\begin{table}[]
\begin{tabular}{||p{0.15\textwidth} p{0.1\textwidth} p{0.1\textwidth} p{0.1\textwidth}|}%{||c|c|c|c|}
\hline
& Mode 7 &  Mode 6 &  Mode 5   \\
\hline\hline
\rule{0pt}{2.2ex} 
$\mu_1$ & 0.28 & 0.41 & 0.46  \\
%\hline
$\mu_2$ & 0.18 & 0.28 & 0.305 \\
%\hline
$\text{QBER}_{Z_{\mu_1}}$ & 1.81 & 4.47 & 2.30 \\
%\hline
$\text{QBER}_{Z_{\mu_2}}$ & 1.92 & 4.28 & 2.09  \\
%\hline
$\text{QBER}_{X_{\mu_1}}$ & 6.24 & 6.75 & 5.89  \\
%\hline
$\text{QBER}_{X_{\mu_2}}$ & 6.02 & 7.59 & 6.71  \\
%\hline
%SKR [kBits/s] & 78 & 17 & 31.8 \\
\hline
\end{tabular}
\caption{Experimental mean photon numbers for the 3-mode multiplexing demonstration as well as QBERs and SKR attained for each mode.}
\label{TAB::3OptimalMu} 
\end{table}
%--- End table
%
%--- Skr figure
\begin{figure}[!h]
\includegraphics[width=0.5\textwidth]{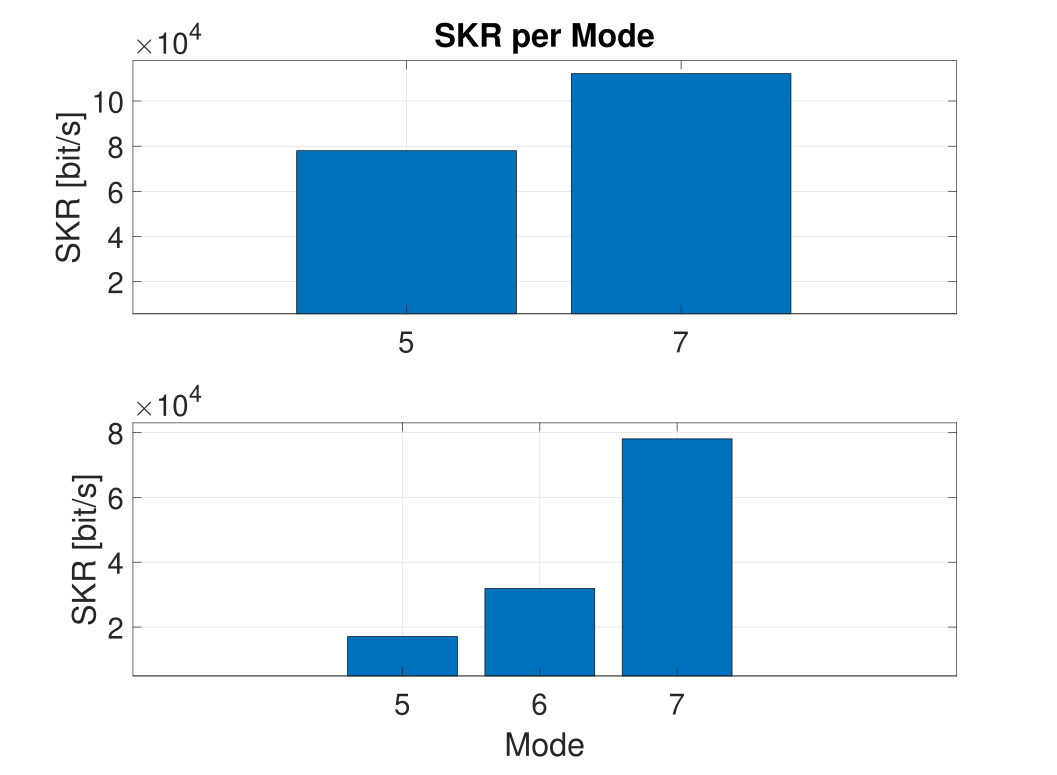}
\caption{Secret key rate experimentally achieved for the 2-mode (top) and 3-mode (bottom) multiplexing.}
\centering
\label{fig:skr}
\end{figure}
%--- End figure
%--- Stability figure
\begin{figure}[!h]
\includegraphics[width=0.5\textwidth]{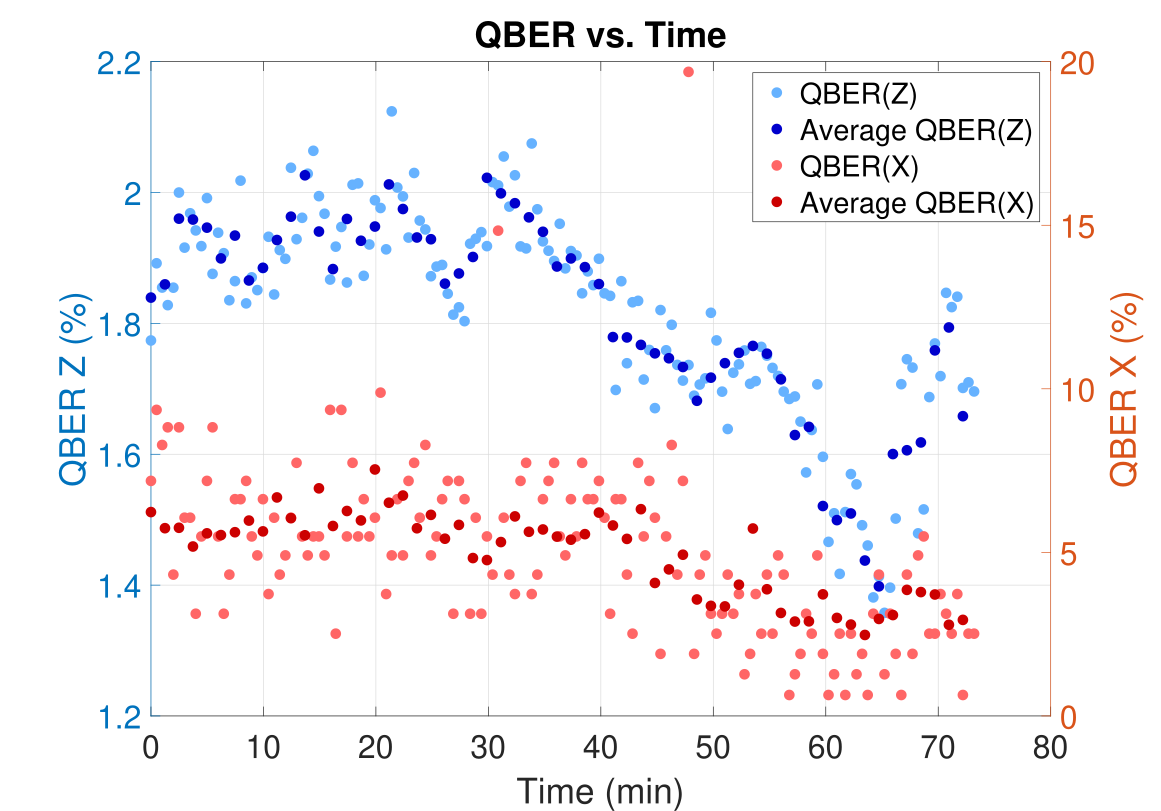}
\caption{Stability of the QKD system. The graph represents variation of QBER measured in the $X$ and $Z$ basis for a duration of approximately 75 minutes. The averaging points represents average QBER obtained every 75s.}
\centering
\label{FIG::LongTermStability_dualAx}
\end{figure}
%--- End figure
%
Since the coupling losses of the demultiplexd light as well as the achievable QBERs were different for each mode due to the different amount of crosstalk, we estimated the optimal mean photon number in each case by using ad-hoc software simulation. These values are reported in Tables \ref{TAB::2OptimalMu} and \ref{TAB::3OptimalMu} for the multiplexing of 2 and 3 QKD signals respectively. The QBER also suffered from background noise due to the optical synchronization scheme, which added $\approx4$ kcounts/s in the $Z$ basis and $\approx200$ kcounts/s in the $X$ basis.
The secret key rate values obtained in each mode 
%together with asymptotic (infinite-size) SKR 
are shown in Figure \ref{fig:skr}. Each point has been measured for 5 minutes.
In addition, in order to test the overall stability of the entire system, we decided to acquire a long measurement of the multiplexing quantum system.
In Figure \ref{FIG::LongTermStability_dualAx}, the long-term stability of key-generation basis QBER, $Q_Z$, and the security check, $Q_X$, are presented. Data acquisition during a course of 75 minutes, while monitoring mode 7 with mean photon number set to $\mu=0.24$, shows a QBER of less than 2\% in the $Z$ basis and 6\% in the $X$ basis. It should be noted that the system is stable for more than an hour and the overall QBER improves during the measurement process. From previous experiences with a similar setup, we can confirm that this effect is due to the phase drift of the heaters in the chip and to the intrinsic stabilization of our measurement apparatus.

\section{Discussion}
In this work we have successfully demonstrated the simultaneous transmission of three different QKD signals multiplexed using OAM fiber modes excited by an integrated photonic chip. Despite the results achieved, multiple improvements should be considered for further deployment of this technology. In particular, improvements on the fabricated silicon photonic chip can be considered. For QKD purposes, for instance, the opto-electronic modules as phase and intensity modulators can be implemented directly on the structure of the chip to form a stand-alone source \cite{sibson2017integrated}. In addition, since the current chip structure only supports one polarization, integrating 2-dimensional grating coupler (2DGC) will enable the chip to propagate orthogonal modes, TM and TE, which will enable new applications such as high dimensional quantum communication. Furthermore, to reduce the losses due to the integrated device, more efficient grating couplers with aluminium mirrors could be added \cite{ding2015efficient}.
An important factor that impairs the fiber modes extinction ratio is the relative phase of each output port on the chip. In the current configuration, the phase can be controlled and manipulated through thermo-optics modules. Carrier depletion modules, as a substitute to thermo-optics modules, can be investigated to increase the extinction ratio. %, but also to increase the repetiton rate.
Indeed, due to the low sensitivity of thermo-optics modules, in our experiment the mode crosstalk increased with the number of active modes, thus increasing the overall QBER and lowering the final secret key rate achievable in the 3-mode test in comparison with 2-mode one. In fact, by exploiting the space division multiplexing (SDM) approach, as demonstrated by D. Bacco and colleagues \cite{bacco2019boosting}, the expected key rate should follow the simple relation $R_{sk}^{tot}= N * R_{sk}$, where $N$ is the number of multiplexed channels considered in the system, and $R_{sk}$ is the secret key rate. In our proof-of-concept, unfortunately, due to the high mode crosstalk this relation could not be demonstrated. 
Finally, a better control over the mode crosstalk would also allow us to excite more than three modes simultaneously, thus increasing the overall secret key rate of the protocol.\\
%--- Future steps
The characterization of the optical modes at the chip output is also a further step to consider. Indeed, the star coupler structure in principle allows for the direct generation of optical modes carrying an OAM. In that case, our integrated device would be regarded as a multiple OAM modes emitter, and it could be implemented in designs for quantum application, for instance as an integrated source of high-dimensional OAM entangled states or for free-space quantum protocols.\\
%--- Final remarks
Summarizing, in our work we have excited, through an integrated silicon chip, OAM fiber modes that have been then used to simultaneously multiplex quantum signals. As a concrete application, we have used the OAM multiplexed signals for demonstrating a QKD protocol based on time-bin encoding, showing the applicability of our system.
These results are of fundamental importance for further developments of OAM related integrated technologies for quantum communication.

%\begin{figure}[h]
%\includegraphics[width=0.5\textwidth]{Figures/QBER_vs_Time_1.eps}
%\caption{Stability of the QKD system. The graph represents variation of QBER measured in the $X$ and $Z$ basis for a duration of approximately 75 minutes. The dashed line represents average QBER obtained every 30s.}
%\centering
%\label{FIG::LongTermStability_singleAx}
%\end{figure}

%\begin{figure}[h]
%\includegraphics[width=0.5\textwidth]{Figures/CrossTalk3D.eps}
%\caption{Mode crosstalk in OAM fiber.}
%\label{Fig::CrossTalk3D}
%\centering
%\end{figure}

%\begin{figure}[h]
%\includegraphics[width=0.5\textwidth]{Figures/SKR_Mode.eps}
%\caption{Achievable secret key rate.}
%\centering
%\end{figure}

%\begin{figure}[h]
%\includegraphics[width=0.5\textwidth]{Figures/SKR_Mode_BarPlot_3.eps}
%\caption{Achievable secret key rate.}
%\centering
%\label{FIG::SKR}
%\end{figure}

\vspace{24pt}
\textbf{Acknowledgment:} We thank S. Ramachandran and P. Gregg for the fiber design, P. Kristensen from OFS-Fitel for the fiber fabrication, I. Vagniluca for the quantum theoretical support and L. S. Rishøj for advices on the fiber alignment and characterization.

\vspace{10pt}
\textbf{Author contribution:} D. B. and D. C. proposed the idea.  M. Z. and Y. L. performed the system experiment. M. Z. realized the quantum system. Y. L. mounted, aligned and characterize the silicon chip and the OAM carrying fibre. D. B. supervised the work. All authors discussed the results and contributed to the writing of the manuscript.

\vspace{10pt}
\textbf{Research funding:} This work is supported by the Center of Excellence SPOC - Silicon Photonics for Optical Communications (ref DNRF123), by the EraNET Cofund Initiatives QuantERA within the European Union’s Horizon 2020 research and innovation program grant agreement No. 731473 (project SQUARE), and by VILLUM FONDEN, QUANPIC (ref. 00025298).

\vspace{10pt}
\textbf{Conflict of interest statement:} The authors declare no conflicts of interest regarding this article.

% --- Bibliography
\bibliographystyle{IEEEtran}
\bibliography{Ref}

\begin{thebibliography}{10}
\providecommand{\url}[1]{#1}
\csname url@samestyle\endcsname
\providecommand{\newblock}{\relax}
\providecommand{\bibinfo}[2]{#2}
\providecommand{\BIBentrySTDinterwordspacing}{\spaceskip=0pt\relax}
\providecommand{\BIBentryALTinterwordstretchfactor}{4}
\providecommand{\BIBentryALTinterwordspacing}{\spaceskip=\fontdimen2\font plus
\BIBentryALTinterwordstretchfactor\fontdimen3\font minus
  \fontdimen4\font\relax}
\providecommand{\BIBforeignlanguage}[2]{{%
\expandafter\ifx\csname l@#1\endcsname\relax
\typeout{** WARNING: IEEEtran.bst: No hyphenation pattern has been}%
\typeout{** loaded for the language `#1'. Using the pattern for}%
\typeout{** the default language instead.}%
\else
\language=\csname l@#1\endcsname
\fi
#2}}
\providecommand{\BIBdecl}{\relax}
\BIBdecl

\bibitem{allen1992orbital}
L.~Allen, M.~W. Beijersbergen, R.~Spreeuw, and J.~Woerdman, ``Orbital angular
  momentum of light and the transformation of laguerre-gaussian laser modes,''
  \emph{Physical review A}, vol.~45, no.~11, p. 8185, 1992.

\bibitem{shen2019optical}
Y.~Shen, X.~Wang, Z.~Xie, C.~Min, X.~Fu, Q.~Liu, M.~Gong, and X.~Yuan,
  ``Optical vortices 30 years on: Oam manipulation from topological charge to
  multiple singularities,'' \emph{Light: Science \& Applications}, vol.~8,
  no.~1, pp. 1--29, 2019.

\bibitem{paterson2001controlled}
L.~Paterson, M.~P. MacDonald, J.~Arlt, W.~Sibbett, P.~Bryant, and K.~Dholakia,
  ``Controlled rotation of optically trapped microscopic particles,''
  \emph{Science}, vol. 292, no. 5518, pp. 912--914, 2001.

\bibitem{padgett2011tweezers}
M.~Padgett and R.~Bowman, ``Tweezers with a twist,'' \emph{Nature photonics},
  vol.~5, no.~6, pp. 343--348, 2011.

\bibitem{erhard2018twisted}
M.~Erhard, R.~Fickler, M.~Krenn, and A.~Zeilinger, ``Twisted photons: new
  quantum perspectives in high dimensions,'' \emph{Light: Science \&
  Applications}, vol.~7, no.~3, pp. 17\,146--17\,146, 2018.

\bibitem{forbes2019quantum}
A.~Forbes and I.~Nape, ``Quantum mechanics with patterns of light: progress in
  high dimensional and multidimensional entanglement with structured light,''
  \emph{AVS Quantum Science}, vol.~1, no.~1, p. 011701, 2019.

\bibitem{willner2015optical}
A.~E. Willner, H.~Huang, Y.~Yan, Y.~Ren, N.~Ahmed, G.~Xie, C.~Bao, L.~Li,
  Y.~Cao, Z.~Zhao \emph{et~al.}, ``Optical communications using orbital angular
  momentum beams,'' \emph{Advances in optics and photonics}, vol.~7, no.~1, pp.
  66--106, 2015.

\bibitem{cozzolino2019orbital}
D.~Cozzolino, D.~Bacco, B.~Da~Lio, K.~Ingerslev, Y.~Ding, K.~Dalgaard,
  P.~Kristensen, M.~Galili, K.~Rottwitt, S.~Ramachandran \emph{et~al.},
  ``Orbital angular momentum states enabling fiber-based high-dimensional
  quantum communication,'' \emph{Physical Review Applied}, vol.~11, no.~6, p.
  064058, 2019.

\bibitem{wang2021high}
Q.-K. Wang, F.-X. Wang, J.~Liu, W.~Chen, Z.-F. Han, A.~Forbes, and J.~Wang,
  ``High-dimensional quantum cryptography with hybrid orbital-angular-momentum
  states through 25 km of ring-core fiber: A proof-of-concept demonstration,''
  \emph{Physical Review Applied}, vol.~15, no.~6, p. 064034, 2021.

\bibitem{bozinovic2013terabit}
N.~Bozinovic, Y.~Yue, Y.~Ren, M.~Tur, P.~Kristensen, H.~Huang, A.~E. Willner,
  and S.~Ramachandran, ``Terabit-scale orbital angular momentum mode division
  multiplexing in fibers,'' \emph{science}, vol. 340, no. 6140, pp. 1545--1548,
  2013.

\bibitem{wang2012terabit}
J.~Wang, J.-Y. Yang, I.~M. Fazal, N.~Ahmed, Y.~Yan, H.~Huang, Y.~Ren, Y.~Yue,
  S.~Dolinar, M.~Tur \emph{et~al.}, ``Terabit free-space data transmission
  employing orbital angular momentum multiplexing,'' \emph{Nature photonics},
  vol.~6, no.~7, pp. 488--496, 2012.

\bibitem{cozzolino2019high}
D.~Cozzolino, B.~Da~Lio, D.~Bacco, and L.~K. Oxenl{\o}we, ``High-dimensional
  quantum communication: benefits, progress, and future challenges,''
  \emph{Advanced Quantum Technologies}, vol.~2, no.~12, p. 1900038, 2019.

\bibitem{ecker2019overcoming}
S.~Ecker, F.~Bouchard, L.~Bulla, F.~Brandt, O.~Kohout, F.~Steinlechner,
  R.~Fickler, M.~Malik, Y.~Guryanova, R.~Ursin \emph{et~al.}, ``Overcoming
  noise in entanglement distribution,'' \emph{Physical Review X}, vol.~9,
  no.~4, p. 041042, 2019.

\bibitem{doerr2011silicon}
C.~R. Doerr, N.~Fontaine, M.~Hirano, T.~Sasaki, L.~Buhl, and P.~Winzer,
  ``Silicon photonic integrated circuit for coupling to a ring-core multimode
  fiber for space-division multiplexing,'' in \emph{European Conference and
  Exposition on Optical Communications}.\hskip 1em plus 0.5em minus 0.4em\relax
  Optical Society of America, 2011, pp. Th--13.

\bibitem{cai2012integrated}
X.~Cai, J.~Wang, M.~J. Strain, B.~Johnson-Morris, J.~Zhu, M.~Sorel, J.~L.
  O’Brien, M.~G. Thompson, and S.~Yu, ``Integrated compact optical vortex
  beam emitters,'' \emph{Science}, vol. 338, no. 6105, pp. 363--366, 2012.

\bibitem{sun2014generating}
J.~Sun, M.~Moresco, G.~Leake, D.~Coolbaugh, and M.~R. Watts, ``Generating and
  identifying optical orbital angular momentum with silicon photonic
  circuits,'' \emph{Optics letters}, vol.~39, no.~20, pp. 5977--5980, 2014.

\bibitem{chen2018mapping}
Y.~Chen, J.~Gao, Z.-Q. Jiao, K.~Sun, W.-G. Shen, L.-F. Qiao, H.~Tang, X.-F.
  Lin, and X.-M. Jin, ``Mapping twisted light into and out of a photonic
  chip,'' \emph{Physical review letters}, vol. 121, no.~23, p. 233602, 2018.

\bibitem{liu2018direct}
J.~Liu, S.-M. Li, L.~Zhu, A.-D. Wang, S.~Chen, C.~Klitis, C.~Du, Q.~Mo,
  M.~Sorel, S.-Y. Yu \emph{et~al.}, ``Direct fiber vector eigenmode
  multiplexing transmission seeded by integrated optical vortex emitters,''
  \emph{Light: Science \& Applications}, vol.~7, no.~3, pp. 17\,148--17\,148,
  2018.

\bibitem{Baumann19}
J.~M. Baumann, K.~Ingerslev, Y.~Ding, L.~H. Frandsen, L.~K. Oxenl{\o}we, and
  T.~Morioka, ``A silicon photonic design concept for a chip-to-fibre orbital
  angular momentum mode-division multiplexer,'' in \emph{2019 Conference on
  Lasers and Electro-Optics Europe and European Quantum Electronics
  Conference}.\hskip 1em plus 0.5em minus 0.4em\relax Optical Society of
  America, 2019, p. pd 1.9.

\bibitem{bacco2019field}
D.~Bacco, I.~Vagniluca, B.~Da~Lio, N.~Biagi, A.~Della~Frera, D.~Calonico,
  C.~Toninelli, F.~S. Cataliotti, M.~Bellini, L.~K. Oxenl{\o}we \emph{et~al.},
  ``Field trial of a three-state quantum key distribution scheme in the
  florence metropolitan area,'' \emph{EPJ Quantum Technology}, vol.~6, no.~1,
  p.~5, 2019.

\bibitem{tebyanian2021practical}
H.~Tebyanian, M.~Zahidy, M.~Avesani, A.~Stanco, P.~Villoresi, and G.~Vallone,
  ``Semi-device independent randomness generation based on quantum state's
  indistinguishability,'' \emph{Quantum Science and Technology}, 2021.

\bibitem{YaoxinLiuECOC21}
Y.~Liu, L.~S. Rish{\o}j, Y.~Ding, Q.~Saudan, L.~K. Oxenl{\o}we, and T.~Morioka,
  ``Orbital angular momentum mode multiplexing and data transmission using a
  silicon photonic integrated mux,'' in \emph{Optical Fiber Communication
  Conference (OFC) 2021}.\hskip 1em plus 0.5em minus 0.4em\relax Optical
  Society of America, 2021, p. F4A.5.

\bibitem{da2021path}
B.~Da~Lio, D.~Cozzolino, N.~Biagi, Y.~Ding, K.~Rottwitt, A.~Zavatta, D.~Bacco,
  and L.~K. Oxenl{\o}we, ``Path-encoded high-dimensional quantum communication
  over a 2-km multicore fiber,'' \emph{npj Quantum Information}, vol.~7, no.~1,
  pp. 1--6, 2021.

\bibitem{sibson2017integrated}
P.~Sibson, J.~E. Kennard, S.~Stanisic, C.~Erven, J.~L. O’Brien, and M.~G.
  Thompson, ``Integrated silicon photonics for high-speed quantum key
  distribution,'' \emph{Optica}, vol.~4, no.~2, pp. 172--177, 2017.

\bibitem{ding2015efficient}
Y.~Ding and K.~Yvind, ``Efficient silicon pic mode multiplexer using grating
  coupler array with aluminum mirror for few-mode fiber,'' in \emph{2015
  Conference on Lasers and Electro-Optics (CLEO)}.\hskip 1em plus 0.5em minus
  0.4em\relax IEEE, 2015, pp. 1--2.

\bibitem{bacco2019boosting}
D.~Bacco, B.~Da~Lio, D.~Cozzolino, F.~Da~Ros, X.~Guo, Y.~Ding, Y.~Sasaki,
  K.~Aikawa, S.~Miki, H.~Terai \emph{et~al.}, ``Boosting the secret key rate in
  a shared quantum and classical fibre communication system,''
  \emph{Communications Physics}, vol.~2, no.~1, pp. 1--8, 2019.

\end{thebibliography}

\end{document}